\documentstyle[emulateapj,epsf]{article}

\newcommand{\bB}{\mbox{\boldmath$B$}}
\newcommand{\bu}{\mbox{\boldmath$u$}}
\newcommand{\degr}{^\circ}

\lefthead{Ogilvie \& Livio}
\righthead{Launching of jets}

\slugcomment{Accepted for publication in the Astrophysical Journal}

\begin{document}

\title{Launching of jets and the vertical structure of accretion disks}

\author{Gordon I. Ogilvie\altaffilmark{1,2,3}}
\and
\author{Mario Livio\altaffilmark{1}}
\altaffiltext{1}{Space Telescope Science Institute, 3700 San Martin
  Drive, Baltimore, MD 21218}
\altaffiltext{2}{Institute of Astronomy,
  University of Cambridge, Madingley Road, Cambridge CB3 0HA, UK}
\altaffiltext{3}{Royal Society University Research Fellow}

\begin{abstract}
  The launching of magnetohydrodynamic outflows from accretion disks
  is considered.  We formulate a model for the local vertical
  structure of a thin disk threaded by a poloidal magnetic field of
  dipolar symmetry.  The model consists of an optically thick disk
  matched to an isothermal atmosphere.  The disk is supposed to be
  turbulent and possesses an effective viscosity and an effective
  magnetic diffusivity.  In the atmosphere, if the magnetic field
  lines are inclined sufficiently to the vertical, a
  magnetocentrifugal outflow is driven and passes through a slow
  magnetosonic point close to the surface.  We determine how the rate
  of mass loss varies with the strength and inclination of the
  magnetic field.  In particular, we find that for disks in which the
  mean poloidal field is sufficiently strong to stabilize the disk
  against the magnetorotational instability, the mass loss rate
  decreases extremely rapidly with increasing field strength, and is
  maximal at an inclination angle of 40--50$^{\circ}$.  For turbulent
  disks with weaker mean fields, the mass loss rate increases
  monotonically with increasing strength and inclination of the field,
  but the solution branch terminates before achieving excessive mass
  loss rates.  Our results suggest that efficient jet launching occurs
  for a limited range of field strengths, and a limited range of
  inclination angles in excess of 30$^{\circ}$.  In addition, we
  determine the direction and rate of radial migration of the poloidal
  magnetic flux, and discuss whether configurations suitable for jet
  launching can be maintained against dissipation.
\end{abstract}


\section{Introduction}

Jets and other outflows are commonly observed from young stellar
objects, interacting binary stars and active galactic nuclei.  It is
generally believed that the outflows are produced by the accretion
disks in these systems.  The widely differing properties of the
central objects in these systems suggest that a mechanism may be at
work that is largely independent of the nature of the central object
(e.g. Livio 1997).

A particularly promising mechanism for the acceleration and
collimation of outflows involves a large-scale poloidal magnetic field
that threads the disk.  The influential model of Blandford \& Payne
(1982) established the significance of the angle of inclination, $i$,
of the poloidal magnetic field lines to the vertical at the surface of
the disk.  When $i>30\degr$, matter that is just above the surface,
being forced to corotate with the foot-point of the field line, is
accelerated outwards along it by the centrifugal force.  At greater
distances, the flow may collimate into a jet by magnetic hoop stresses
or poloidal collimation.  A clear review of the physics of
magnetocentrifugally driven outflows has been given by Spruit (1996).

Several groups have performed axisymmetric numerical simulations of
outflows using this mechanism (e.g.~Ustyugova et al. 1995; Romanova
et~al.\ 1997; Ustyugova et~al.\ 1999; Ouyed \& Pudritz 1997a, 1997b,
1999; Krasnopolsky, Li, \& Blandford 1999).  Like the model of
Blandford \& Payne (1982), these calculations do not resolve the
vertical structure of the disk but treat it as a boundary surface that
loads mass at a specified rate on to the magnetic field lines.  While
these simulations have convincingly demonstrated some aspects of the
magnetocentrifugal model of jet production, several fundamental issues
remain to be resolved.  In this paper, we will attempt to address two
of these questions.  First, what determines the rate of mass loss in
the outflow?  How does this depend on the strength and inclination of
the magnetic field, or on other properties of the disk?  Second, what
is the long-term evolution of a large-scale magnetic field in a disk?
Is it possible to assemble a magnetic configuration suitable for jet
launching, or to maintain it against dissipation?  A proper
understanding of these issues is essential if we are to explain the
conditions that regulate the production of astrophysical jets.

The magnetocentrifugal mechanism can explain the acceleration of
outflows even when the matter is `cold' in the sense that the
temperature is much less than the virial temperature or escape
temperature.  However, as mentioned by Blandford \& Payne (1982) and
calculated in detail by Ogilvie (1997) and by Ogilvie \& Livio (1998;
hereafter Paper~I), some amount of thermal assistance is still
required to overcome the potential barrier between the surface and the
slow magnetosonic point (`sonic point').  The barrier occurs because
the field lines are not straight inside the disk and the angular
velocity deviates slightly from the Keplerian value because of the
radial Lorentz force associated with the bending of the field lines
(Shu 1991; Wardle \& K\"onigl 1993).  A proper calculation of the
vertical structure of the disk, including these effects, is therefore
required in order to determine the height of the potential barrier and
the rate of mass loss in the outflow.  Indeed, it is obvious from the
well-known properties of transonic outflows that the mass loss rate
depends on the physical conditions below the sonic point, and that the
vertical structure must therefore be resolved.

In Paper~I we carried out such a calculation for disks that are rather
strongly magnetized in the sense that the magnetorotational
instability, which leads to turbulence in accretion disks (Balbus \&
Hawley 1998), is suppressed or nearly so.  We assumed that the
magnetic field enforces strict isorotation and showed that the
potential barrier increases very steeply as the field is made stronger
and the disk becomes more sub-Keplerian.  This effect would suppress
outflows from strongly magnetized disks unless an additional source of
energy, such as coronal heating, were present (in accord with the
suggestion made by Livio 1997).

A disadvantage of the calculation in Paper~I is that no explanation
was given for the source of the effective viscosity of a disk in which
the magnetorotational instability is suppressed.  The solutions did
not extend to field strengths much below the stability boundary.  When
the field is weaker than this, it cannot be expected to enforce
isorotation, and the model requires some modification.  In addition,
the effect of the turbulence on the mean magnetic field should be
modeled, most simply through the introduction of an effective
magnetic diffusivity.

However, the presence of a turbulent diffusivity may cause problems
for the magnetocentrifugal mechanism.  The analysis by Lubow,
Papaloizou, \& Pringle (1994a) suggests that, if the effective
magnetic Prandtl number of the disk is of order unity, as might be
expected, it is impossible to sustain a steady configuration with a
significantly inclined field in a thin disk (see also Heyvaerts,
Priest, \& Bardou 1996; Reyes-Ruiz \& Stepinski 1996).  Although the
accretion flow tends to drag magnetic flux inwards, the turbulent
diffusivity expels flux faster if the inclination is large.  The
inclination angle in a steady state is then expected to be comparable
to the angular semi-thickness $H/r$ of the disk.  This constraint can
be avoided if the accretion flow is due primarily to the loss of
angular momentum in an outflow, and might also be relieved by a dynamo
operating in the disk if special conditions are met (Campbell,
Papaloizou, \& Agapitou 1998).

The purpose of the present paper is to explore a model for the
vertical structure of magnetized disks.  We will extend the analysis
of Paper~I to allow for the possibility of more weakly magnetized
disks in which there is turbulence, and strict isorotation does not
hold.  We will also give a better treatment of the matching between
the disk and the atmosphere by applying photospheric boundary
conditions, allowing us to calculate the mass loss rate in the outflow
explicitly.  At the same time, we will solve for the rate of dragging
of magnetic flux and thereby refine the previous estimates which may
have been based on oversimple arguments.

Related calculations of the vertical structure of magnetized disks,
and of the disk-jet connection, have been made by K\"onigl (1989),
Wardle \& K\"onigl (1993), Li (1995, 1996), Ferreira (1997), Campbell
(1999), Casse \& Ferreira (2000), and Shalybkov \& R\"udiger (2000).
Our method and results are significantly different from all previous
calculations and comparisons of the relevant issues will be made
towards the end of this paper.

\section{Context of the present calculation}

To set this calculation in its proper context, we consider here some
of the wider issues relating to magnetized accretion disks and
outflows.

For a thin disk without a large-scale magnetic field, there is a clear
division of the physical problem into two aspects.  First, a model is
required for the local vertical structure at any given radius $r$ and
time $t$.  Such a model takes the surface mass density $\Sigma$ as a
parameter and predicts quantities such as the vertically integrated
viscous stress ${\cal G}$.  The vertical structure may be assumed to
be stationary on the dynamical time-scale.  Second, the
one-dimensional conservation equations for mass and angular momentum
determine how $\Sigma(r,t)$ evolves on the viscous time-scale, given
the local relation between ${\cal G}$ and $\Sigma$ (e.g. Lynden-Bell
\& Pringle 1974).

For a disk with a large-scale poloidal magnetic field but no outflows
(e.g. a polytrope with a force-free or vacuum exterior), the situation
is similar but more complicated.  Now the local vertical structure at
a given radius depends not only on $\Sigma$ but also on the vertical
magnetic field $B_z$ and the inclination angle $i$.  In addition to an
evolutionary equation for $\Sigma$, there is a one-dimensional
conservation equation for magnetic flux, which determines how the flux
function $\psi(r,t)$ evolves on the viscous time-scale.  $B_z$ is
simply related to the radial derivative of $\psi$.  Finally, the
force-free magnetic equilibrium in the exterior of the disk leads to a
global relation between $\psi$ and the inclination angle $i$.  The
full problem therefore involves an integro-differential equation.
This has been considered by Lubow et al. (1994a), although they did not
examine the effect of the magnetic field on the vertical disk
structure.

When outflows occur along the magnetic field lines, further couplings
exist.  The exterior magnetic field is modified from the force-free
solution by the inertial forces associated with the outflow.  This
effect depends on the amount of mass loading, and is therefore coupled
to the launching problem.  The exterior field is no longer purely
poloidal but becomes significantly twisted beyond the Alfv\'en
surface, although non-axisymmetric (e.g. kink) instabilities may set
in here (e.g. Spruit, Foglizzo, \& Stehle 1997).  This part of the
solution is established on the Alfv\'en travel time to the Alfv\'en
surface, which is comparable to the dynamical time-scale of the disk.
Finally, the mass loss and especially the angular momentum loss in the
outflow feed back into the evolutionary equations for the disk.

The full problem therefore involves three aspects: the local vertical
structure of the disk and the launching problem; the global structure
of the exterior magnetic field and the dynamics of the outflow along
those field lines; and the evolution of mass, angular momentum, and
magnetic flux on a longer time-scale.  The first aspect is the subject
of this paper, while the second aspect is well described by some of
the numerical simulations referred to above.  The third aspect has
been considered implicitly in steady models such as that of Casse \&
Ferreira (2000), but the development of a complete evolutionary scheme
for mass, angular momentum and magnetic flux in the general,
non-steady case remains a challenge (see Lovelace, Newman, \& Romanova
1997 for a simplified version of such a scheme).\footnote{There have
  been many calculations simulating the time-dependent development of
  a jet from a numerically resolved disc.  For example, Kudoh,
  Matsumoto, \& Shibata (1998) set up a thick torus with a
  non-rotating corona, and introduced a vertical magnetic field.
  Owing to the lack of equilibrium in the initial condition, a rapid
  adjustment occurs, and Kudoh et al. followed the evolution for only
  a single orbit or so.  Unfortunately there is little reason to
  expect the properties of such transient phenomena to agree with
  quasi-steady models of jets from thin discs.}

There has been some confusion in the literature regarding the
interplay between these aspects of the problem.  For example, studies
of the launching problem have often imposed the condition that the
poloidal magnetic flux should not migrate radially (e.g. K\"onigl
1989; Li 1995).  That is, there should be an instantaneous balance
between inward advection of flux by the accretion flow, and outward
transport due to turbulent, Ohmic, or ambipolar diffusion.  This may
not always be appropriate because the migration occurs on the viscous
time-scale, whereas the solution of the launching problem need only be
stationary on the dynamical time-scale.  In Wardle \& K\"onigl (1993)
the rate of flux migration was treated as a free parameter.

Moreover, some of these studies appear to predict the value of the
toroidal magnetic field at the surface of the disk, $B_{\phi{\rm s}}$,
even when the dynamics of the outflow beyond the sonic point has not
been considered (e.g. Campbell 1999) or when there is no outflow (e.g.
Shalybkov \& R\"udiger 2000).  This is unsatisfactory because
$B_{\phi{\rm s}}$ determines the external magnetic torque acting on
the disk, which certainly depends on the dynamics of the outflow in
the trans-Alfv\'enic region.  These incorrectly specified studies may
be motivated by the fact that the model of Blandford \& Payne (1982)
appears to require $B_{\phi{\rm s}}$ to be determined by the disk
(through their parameter $\lambda$).  However, this is misleading and
may be attributed to the fact that the Blandford \& Payne model is
underconstrained because it is missing a boundary condition at large
distances from the disk (cf.\ Ostriker 1997).  Indeed, solutions of
their model generically behave unphysically at large distances without
having passed through the modified fast magnetosonic point. In
contrast, Krasnopolsky et~al.\ (1999) have given a clear discussion of
which quantities may or may not be specified as boundary conditions
when treating the disk as a boundary surface.  In their model
$B_{\phi{\rm s}}$ is determined by the outflow, not by the disk.

In our local study, therefore, we will not impose the constraint of
zero flux migration; rather, we will impose the value of $B_{\phi{\rm
    s}}$ and determine the rate of flux migration.  Usually we will
specify $B_{\phi{\rm s}}=0$, meaning that the outflow is absent or
exerts only a weak torque.  Efficient magnetocentrifugal outflows are
expected to have $|B_{\phi{\rm s}}|\ll|B_z|$, with the toroidal field
becoming comparable to the poloidal component only at greater
distances from the source, where the Alfv\'en surface is located
(Spruit 1996).

\section{The evolution of magnetic flux}

As noted above, an important investigation of the evolution of the
poloidal magnetic flux was made by Lubow et~al.\ (1994a), who concluded
that, if the disk is turbulent, with an effective magnetic Prandtl
number of order unity, the accretion flow will be almost entirely
ineffective in dragging in magnetic flux.  This can be explained by
noting that the effective magnetic Reynolds number of the accretion
flow, based on the disk thickness $H$, is small, of order $H/r$
(Heyvaerts et~al.\ 1996).  Similar arguments suggest that a
configuration in which the field lines are bent significantly from the
vertical (e.g.\ to achieve $i>30\degr$) cannot be sustained on the
viscous time-scale.

These arguments are based on a kinematic analysis of the magnetic
induction equation in which the radial velocity and magnetic
diffusivity are prescribed quantities.  They also depend on a simple
order-of-magnitude treatment of the vertical structure of the magnetic
field.  In this paper we will present a numerical treatment of a set
of equations, including the induction equation, in which the radial
velocity is self-consistently determined.  This is important because,
in a jet-launching configuration, the magnetic field must become
dynamically dominant above some height and will then control the
radial velocity.  We will find results that differ significantly from
the estimates of Lubow et al. (1994) under some circumstances.  Before
presenting the numerical model, we therefore reconsider the problem of
the induction equation from an analytical viewpoint.

We assume that the mean poloidal magnetic field is axisymmetric and
may be described by a flux function $\psi(r,z,t)$ such that
\begin{equation}
  B_r=-{{1}\over{r}}{{\partial\psi}\over{\partial z}},\qquad
  B_z={{1}\over{r}}{{\partial\psi}\over{\partial r}}.
\end{equation}
For a thin disk containing a bending poloidal field of dipolar
symmetry, the flux function has the form (Ogilvie 1997)
\begin{equation}
  \psi=\psi_0(r,t)+\psi_1(r,z,t),
\end{equation}
where $|\psi_1|\ll|\psi_0|$.  Then $B_r$ and $B_z$ are comparable in
magnitude if $|\psi_1/\psi_0|=O(H/r)$.  The flux content of the disk
is determined essentially by $\psi_0$, while $\psi_1$ allows for
bending of the field lines within the disk.

We assume that the mean magnetic field in a turbulent disk may be
treated within the framework of mean-field electrodynamics (e.g.\ 
Moffatt 1978). In the absence of a mean-field dynamo effect (or in the
absence of a toroidal field), the flux function then satisfies the
mean induction equation,
\begin{equation}
  {{\partial\psi}\over{\partial t}}+\bu\!\cdot\!\nabla\psi=
  \eta r^2\nabla\!\cdot\left({{1}\over{r^2}}\nabla\psi\right),
\end{equation}
where $\bu$ is the mean velocity and $\eta$ the turbulent magnetic
diffusivity.  In a thin disk the dominant terms are
\begin{equation}
  {{\partial\psi_0}\over{\partial t}}+
  u_r{{\partial\psi_0}\over{\partial r}}=
  \eta r{{\partial}\over{\partial r}}
  \left({{1}\over{r}}{{\partial\psi_0}\over{\partial r}}\right)+
  \eta{{\partial^2\psi_1}\over{\partial z^2}}.
\end{equation}
The neglected terms are all smaller than those retained because
$|\psi_1|\ll|\psi_0|$ and $|u_r|\gg|u_z|$.  The first term on the
right-hand side is also sometimes neglected (e.g. Lubow et al. 1994a).

Although this equation appears to be an evolutionary equation for
$\psi_0$ it should be observed that the equation is defined for all
values of $z$ whereas $\psi_0$ is a function of $r$ and $t$ only.
Moreover, the equation is not closed because of the term involving
$\psi_1$.  Following Lubow et al. (1994a), we might attempt to close
the equation by vertical averaging.  Conventionally in accretion-disk
theory one uses density-weighted averages $\bar u_r$ and $\bar\eta$
defined by
\begin{equation}
  \Sigma\bar u_r=\int\rho u_r\,{\rm d}z,\qquad
  \Sigma\bar\eta=\int\rho\eta\,{\rm d}z,
\end{equation}
where $\rho$ is the density and
\begin{equation}
  \Sigma=\int\rho\,{\rm d}z
\end{equation}
is the surface density, the integrals being over the full vertical
extent of the disk.  The density-weighted average $\bar u_r$ is
particularly appropriate because it is directly related to the mass
accretion rate.  The vertical average of the induction equation is
then
\begin{equation}
  {{\partial\psi_0}\over{\partial t}}+
  \bar u_r{{\partial\psi_0}\over{\partial r}}=
  \bar\eta r{{\partial}\over{\partial r}}
  \left({{1}\over{r}}{{\partial\psi_0}\over{\partial r}}\right)-
  {{r}\over{\Sigma}}\int\rho\eta{{\partial B_r}\over{\partial z}}\,{\rm d}z.
\end{equation}
Lubow et al. proceeded by approximating the last term as
$-(r/H)\bar\eta B_{r{\rm s}}$.  This closes the equation because
$B_{r{\rm s}}$, which is proportional to the vertically integrated
toroidal electric current, can be globally related to the flux
function by considering the force balance exterior to the disk.  If
the exterior region is treated either as an insulating vacuum, or as a
force-free medium with vanishing $B_\phi$, the exterior poloidal field
is potential.  This leads to a global relation of the form
\begin{equation}
  B_{r{\rm s}}={\cal L}^{-1}\psi_0,
\end{equation}
where ${\cal L}$ is a certain linear integral operator (see also
Ogilvie 1997).

This order-of-magnitude estimate of the final term appears reasonable
if the field lines bend mainly in the densest layers of the disc near
the midplane.  However, the term might be significantly overestimated
if the bending occurs mainly in the upper layers of the disk where
$\rho\eta$ is likely to be considerably smaller.  Another way of
averaging the induction equation reinforces this concern.  If the
equation is divided by $\eta$ and integrated with respect to $z$, we
obtain
\begin{equation}
  {{\partial\psi_0}\over{\partial t}}+
  u_*{{\partial\psi_0}\over{\partial r}}=
  \eta_*r{{\partial}\over{\partial r}}
  \left({{1}\over{r}}{{\partial\psi_0}\over{\partial r}}\right)-
  QB_{r{\rm s}},
  \label{dpsi0}
\end{equation}
where
\begin{equation}
  u_*(r,t)=\int{{u_r}\over{\eta}}\,{\rm d}z\bigg/
  \int{{1}\over{\eta}}\,{\rm d}z,
\end{equation}
\begin{equation}
  \eta_*(r,t)=\int{\rm d}z\bigg/\int{{1}\over{\eta}}\,{\rm d}z,
\end{equation}
and
\begin{equation}
  Q(r,t)=2r\bigg/\int{{1}\over{\eta}}\,{\rm d}z.
\end{equation}
Now the precise shape of the field line has truly been eliminated in
favor of $B_{r{\rm s}}$.  This therefore appears to be the `correct'
way of averaging the equation.  However, we are now faced with
vertical averages of $u_r$ and $\eta$ weighted by $1/\eta$, not by
$\rho$.  Even if $\eta$ is independent of $z$, these could be
significantly different from the density-weighted averages, leading to
different conclusions about flux dragging.

Furthermore, as mentioned above, the presence of the magnetic field
will in general change the vertical profiles of $u_r$ and $\eta$.  If
the field is extremely weak so that the induction equation may be
treated `kinematically', then the $1/\eta$ averaging method is surely
the correct one.  However, the vertical profile of $u_r$, in
particular, is determined by subtle effects (e.g.  Kley \& Lin 1992)
and is easily distorted by even weak Lorentz forces.  In this case we
cannot obtain a closed equation because $u_*$ cannot be determined
from a knowledge of the mass accretion rate.  The essential difficulty
is that, while $\bar u_r$ is the relevant mean velocity for mass
accretion, $u_*$ is the appropriate mean for flux accretion.  These
could be significantly disparate and could even differ in sign.  The
above discussion shows that the only solution to this problem is to
solve explicitly for the vertical structure of the disk including the
physics that determines the vertical profile of the radial velocity.

\section{Mathematical model}

We now consider the set of equations governing the local vertical
structure of a disk containing a mean poloidal magnetic field.  The
equations of Paper~I will be augmented by the inclusion of additional
physical effects.

\subsection{Basic equations for a thin disk}

The angular velocity is written as
\begin{equation}
  \Omega=\Omega_0(r)+\Omega_1(r,z,t),
\end{equation}
where
\begin{equation}
  \Omega_0=\left({{GM}\over{r^3}}\right)^{1/2}
\end{equation}
is the Keplerian value, and $|\Omega_1|\ll\Omega_0$.

The flux function is written as
\begin{equation}
  \psi=\psi_0(r,t)+\psi_1(r,z,t),
\end{equation}
where $|\psi_1|\ll|\psi_0|$, and we approximate
\begin{equation}
  B_r=-{{1}\over{r}}{{\partial\psi_1}\over{\partial z}},\qquad
  B_z\approx{{1}\over{r}}{{\partial\psi_0}\over{\partial r}},
\end{equation}
so that $B_z$ is independent of $z$.

The required equations may be approximated as follows.  The radial
component of the equation of motion is
\begin{equation}
  -2\rho r\Omega_0\Omega_1=
  {{B_z}\over{\mu_0}}{{\partial B_r}\over{\partial z}},
  \label{motion_r}
\end{equation}
where $\mu_0$ is the permeability of free space.  The azimuthal
component is
\begin{equation}
  {{\rho u_r}\over{r}}{{d}\over{dr}}(r^2\Omega_0)=
  {{B_z}\over{\mu_0}}{{\partial B_\phi}\over{\partial z}}+
  {{1}\over{r^2}}{{\partial}\over{\partial r}}
  \left(\rho\nu r^3{{d\Omega_0}\over{dr}}\right),
  \label{motion_phi}
\end{equation}
where $\nu$ is the kinematic viscosity.  The vertical component is
\begin{equation}
  0=-\rho\Omega_0^2z-
  {{\partial}\over{\partial z}}
  \left(p+{{B_r^2}\over{2\mu_0}}+{{B_\phi^2}\over{2\mu_0}}\right),
  \label{motion_z}
\end{equation}
where $p$ is the pressure.  The poloidal part of the induction
equation is
\begin{equation}
  {{\partial\psi_0}\over{\partial t}}+ru_r B_z=
  r\eta\left({{\partial B_z}\over{\partial r}}-
  {{\partial B_r}\over{\partial z}}\right).
  \label{induction_rz}
\end{equation}
The toroidal part is
\begin{equation}
  0=r\left(B_r{{d\Omega_0}\over{dr}}+
  B_z{{\partial\Omega_1}\over{\partial z}}\right)+
  {{\partial}\over{\partial z}}
  \left(\eta{{\partial B_\phi}\over{\partial z}}\right).
  \label{induction_phi}
\end{equation}
The energy equation is
\begin{equation}
  {{\partial F}\over{\partial z}}=
  \rho\nu\left(r{{d\Omega_0}\over{dr}}\right)^2+
  {{\eta}\over{\mu_0}}
  \left[\left({{\partial B_r}\over{\partial z}}\right)^2+
  \left({{\partial B_\phi}\over{\partial z}}\right)^2\right],
  \label{energy}
\end{equation}
where
\begin{equation}
  F=-{{16\sigma T^3}\over{3\kappa\rho}}{{\partial T}\over{\partial z}}
  \label{f}
\end{equation}
is the radiative energy flux, with $\sigma$ the Stefan-Boltzmann
constant, $T$ the temperature, and $\kappa$ the Rosseland mean
opacity.

These equations must be supplemented by constitutive relations
specifying the equation of state, the opacity, the viscosity, and the
magnetic diffusivity.  We will adopt the ideal-gas equation of state,
\begin{equation}
  p={{k\rho T}\over{\mu m_{\rm H}}},
\end{equation}
(where $k$ is the Boltzmann constant, $\mu$ the mean molecular mass,
and $m_{\rm H}$ the mass of the hydrogen atom), and a generic
power-law opacity,
\begin{equation}
  \kappa=C_\kappa\rho^xT^y,
\end{equation}
where $C_\kappa$, $x$, and $y$ are constants.  This includes the cases
of Thomson scattering opacity ($x=y=0$) and Kramers opacity ($x=1$,
$y=-7/2$).

\subsection{Turbulent viscosity and magnetic diffusivity}

The viscosity is less certain, and we will adopt the standard
prescription
\begin{equation}
  \rho\nu={{\alpha p}\over{\Omega_0}},
\end{equation}
where $\alpha$ is a dimensionless constant.  For the magnetic
diffusivity, we assume that the magnetic Prandtl number,
\begin{equation}
  {\rm Pm}={{\nu}\over{\eta}},
\end{equation}
is constant.

We will further assume either that $\alpha$ is a fixed parameter
(`{\it fixed alpha hypothesis}'), or that $\alpha$ can adjust so as to
keep the equilibrium at marginal magnetorotational stability, as
explained in Section~4.8 below (`{\it marginal stability
  hypothesis}').

\subsection{Neglected terms}

Several terms in the equations have been omitted on the grounds that
the disk is thin and the solution should be stationary on the
dynamical time-scale (although not necessarily on the viscous
time-scale).  The radial pressure gradient, the vertical variation of
radial gravity, and the inertial terms associated with the meridional
flow, have been neglected as usual.  However, enough terms have been
retained to determine the profile of radial velocity in the absence of
a magnetic field.

Generally, $\Omega_1$ has been neglected relative to $\Omega_0$, and
$\psi_1$ relative to $\psi_0$, except where physically essential.  It
has also been assumed that the viscosity does not act on the shear
components $\partial u_r/\partial z$ and $r\partial\Omega_1/\partial
z$.  Such terms are expected to be relatively unimportant in most
cases of interest, and it was found that the inclusion of these terms
increases the order of the differential system and makes it difficult
to obtain a solution.  It is especially unclear how to connect a
viscous disk to an inviscid atmosphere, when these terms are included,
without introducing an artificial boundary layer.

We did not include a dynamo alpha-effect in the equations for the mean
magnetic field.  Although there is no difficulty in principle in doing
so, this effect is even less well understood than the turbulent
viscosity and magnetic diffusivity, and we chose to avoid this
additional complication in the present study.

Finally, in the energy equation (\ref{energy}) it has been assumed
that the heat generated in each annulus is radiated away locally and
not advected through the disk.

\subsection{Local and non-local effects}

The equations of Section~4.1 describe the local vertical structure of
an accretion disk with a mean poloidal magnetic field.  When
supplemented with appropriate boundary conditions, as described below,
they constitute a problem similar in type to a stellar structure
calculation, although the details are of course very different.
Obviously it is possible in principle to include a more detailed
equation of state, accurate opacity tables, a more sophisticated
approach to radiative transfer, and further possibilities such as
convective energy transport.  However, the principal uncertainties
concern the viscosity and magnetic diffusivity.

In fact, these equations are not strictly local in radius and time
because some radial derivatives and one time-derivative remain.  The
case of $d\Omega_0/dr=-3\Omega_0/2r$ is trivial.  The terms
$\partial\psi_0/\partial t$ and $\partial B_z/\partial r$ in the
poloidal part of the induction equation are retained in order to
determine whether the magnetic flux migrates inwards
($\partial\psi_0/\partial t>0$) or outwards as a result of the local
disk solution.  The gradient $\partial B_z/\partial r$ can affect this
result because it contributes to the diffusion of flux.  Therefore
$\partial B_z/\partial r$ appears as a parameter of the local model
and $\partial\psi_0/\partial t$ as an eigenvalue; note that both are
independent of $z$.

The case of $\partial(\rho\nu)/\partial r$ is more problematic.  This
viscous term is retained in the angular momentum equation because it
partially determines the radial velocity, which in turn causes radial
advection of magnetic flux and also affects the shape of the field
lines, at least when the field is weak.  This term may be approximated
by arguing that
\begin{equation}
  \rho\nu\approx{{\bar\nu\Sigma}\over{H}}f\left({{z}\over{H}}\right)
\end{equation}
in the neighborhood of the radius under consideration, where
\begin{equation}
  \bar\nu\Sigma=\int_{-H}^H\rho\nu\,dz
\end{equation}
is the vertically integrated dynamic viscosity, $H$ the
semi-thickness, and $f$ an undetermined dimensionless function.  Under
this assumption,
\begin{equation}
  {{\partial\ln(\rho\nu)}\over{\partial\ln r}}=
  {{\partial\ln(\bar\nu\Sigma)}\over{\partial\ln r}}-
  \left[1+{{\partial\ln(\rho\nu)}\over{\partial\ln z}}\right]
  {{\partial\ln H}\over{\partial\ln r}}.
\end{equation}
The vertical derivative is available as part of the local solution,
while $\partial\ln(\bar\nu\Sigma)/\partial\ln r$ and $\partial\ln
H/\partial\ln r$ appear as additional dimensionless parameters, which
can be estimated from the well-known behavior of the steady,
non-magnetized solution (Shakura \& Sunyaev 1973).

In the limit of a non-magnetized disk, this prescription predicts the
radial velocity in the disk as
\begin{equation}
  u_r=-{{3\nu}\over{2r}}
  \left\{1+2{{\partial\ln(\bar\nu\Sigma)}\over{\partial\ln r}}-
  2\left[1+{{\partial\ln(\rho\nu)}\over{\partial\ln z}}\right]
  {{\partial\ln H}\over{\partial\ln r}}\right\}.
  \label{ur}
\end{equation}

\subsection{Boundary conditions}

We have derived a sixth-order system of nonlinear ordinary
differential equations (ODEs) in $z$.  The dependent variables may be
taken as $p$, $T$, $F$, $\Omega_1$, $B_r$, and $B_\phi$.  The
variables $\rho$ and $u_r$ are determined algebraically from these.

The solution should be symmetric about the mid-plane, with
\begin{equation}
  F=B_r=B_\phi=0
\end{equation}
at $z=0$.

The solution extends up to a photospheric surface $z=H$ at which it is
matched to a simple atmospheric model, discussed below.  The
photospheric boundary conditions are
\begin{equation}
  F_{\rm s}=\sigma T_{\rm s}^4
\end{equation}
and
\begin{equation}
  \tau_{\rm s}=\int_H^\infty\kappa\rho\,dz={{2}\over{3}},
  \label{tau}
\end{equation}
where the subscript `s' denotes a surface value.  We also have
\begin{equation}
  B_r=B_{r{\rm s}},\qquad
  B_\phi=B_{\phi{\rm s}}
\end{equation}
there, where $B_{r{\rm s}}$ and $B_{\phi{\rm s}}$ are assigned values
which, physically, are determined by the solution of the global
exterior outflow problem (not considered here).

With these boundary conditions the equation of angular momentum
conservation, obtained from a vertical integration of equation
(\ref{motion_phi}), has the form
\begin{equation}
  -{{\dot M}\over{2\pi r^2}}{{d}\over{dr}}(r^2\Omega_0)=
  {{2B_zB_{\phi{\rm s}}}\over{\mu_0}}+
  {{1}\over{r^2}}{{\partial}\over{\partial r}}
  \left(\bar\nu\Sigma r^3{{d\Omega_0}\over{dr}}\right),
  \label{am}
\end{equation}
where
\begin{equation}
  \dot M=-2\pi r\int_{-H}^H\rho u_r\,dz
\end{equation}
is the mass accretion rate, not necessarily constant.  This shows the
contributions to angular momentum transport from magnetic and viscous
torques.

The solution is determined as follows.  One guesses the values of the
quantities $H$, $F_{\rm s}$, and $\Omega_{1{\rm s}}$.  This fixes the
atmospheric model and determines $p_{\rm s}$ and $T_{\rm s}$.  The
quantities $B_{r{\rm s}}$ and $B_{\phi{\rm s}}$ are given.  One must
further guess $\partial\psi_0/\partial t$ to start the downward
integration. The three guessed quantities should be adjusted to match
the three symmetry conditions on $z=0$.  The main parameters of the
model are then $\Omega_0$, $\Sigma$, $B_z$, $B_{r{\rm s}}$,
$B_{\phi{\rm s}}$, $\alpha$, and ${\rm Pm}$.  The quantity
$\partial\psi_0/\partial t$ is to be determined as an eigenvalue.

\subsection{Atmospheric model}

In the simplest atmospheric model, $T$, $F$, $B_r$, and $B_\phi$ are
independent of $z$ between the photosphere and the sonic point, with
\begin{eqnarray}
  T&=&T_{\rm s}=T_{\rm eff},\\
  F&=&F_{\rm s}=\sigma T_{\rm s}^4,\\
  B_r&=&B_{r{\rm s}}=B_z\tan i,\\
  B_\phi&=&B_{\phi{\rm s}},
\end{eqnarray}
where $i$ is the inclination of the field lines to the vertical.  The
constancy of $F$ relies on there being little or no dissipation in the
atmosphere, while the constancy of $\bB$ relies on the atmosphere
being magnetically dominated so that field line bending cannot be
supported.  That is, the plasma beta based on the poloidal magnetic
field strength,
\begin{equation}
  \beta={{2\mu_0p}\over{B_r^2+B_z^2}},
\end{equation}
should be less than unity.

In the atmosphere, all diffusive terms in the equations are neglected.
If $\partial\psi_0/\partial t\neq0$, there must be a slow drift
orthogonal to the poloidal field lines, but otherwise the flow is
constrained to follow the field.  We know from Paper~I that there will
be a transonic outflow if $i>30\degr$, otherwise a modified
hydrostatic atmosphere.

As in Paper~I, the angular velocity in the atmosphere is determined by
isorotation,
\begin{equation}
  \Omega_1=\Omega_{1{\rm s}}+{{3\Omega_0}\over{2r}}(z-H)\tan i.
\end{equation}
Again, this assumes that the atmosphere is magnetically dominated.
The centrifugal-gravitational potential for matter constrained to
follow the field is
\begin{equation}
  \Phi^{\rm cg}=-{{1}\over{2}}(3\tan^2i-1)\Omega_0^2(z-z_{\rm sonic})^2,
\end{equation}
where
\begin{equation}
  z_{\rm sonic}=H\left[1+(3\tan^2i-1)^{-1}
  \left(1-2{{\Omega_{1{\rm s}}r}\over{\Omega_0H}}\tan i\right)\right]
\end{equation}
is the height of the sonic point in the case $i>30\degr$.

The density scale-height $h_{\rm s}$ at the photosphere is given by
\begin{equation}
  {{c_{\rm s}^2}\over{h_{\rm s}}}=
  \left({{\partial\Phi^{\rm cg}}\over{\partial z}}\right)_{\rm s}
  =\Omega_0^2H\left(1-2{{\Omega_{1{\rm s}}r}\over{\Omega_0H}}\tan i\right),
\end{equation}
where $c_{\rm s}$ is the isothermal sound speed, given by
\begin{equation}
  c_{\rm s}^2={{kT_{\rm s}}\over{\mu m_{\rm H}}}.
\end{equation}
To a good approximation, the photospheric boundary condition
(\ref{tau}) equates to
\begin{equation}
  {{2}\over{3}}=C_\kappa\rho_{\rm s}^{1+x}T_{\rm s}^y\int_0^\infty
  \exp\left[-(1+x)\left({{z-H}\over{h_{\rm s}}}\right)\right]\,d(z-H).
\end{equation}
This may be rearranged into the form
\begin{eqnarray}
  \rho_{\rm s}^{1+x}T_{\rm s}^{1+y}&=&
  \left({{1+x}\over{8\sigma}}\right)\left({{\mu m_{\rm H}}\over{k}}\right)
  \left({{16\sigma}\over{3C_\kappa}}\right)\Omega_0^2H
  \nonumber\\
  &&\qquad\times
  \left(1-2{{\Omega_{1{\rm s}}r}\over{\Omega_0H}}\tan i\right).
\end{eqnarray}

Finally, in the case of a transonic outflow, the Mach number at the
surface is determined from the equation (see Paper~I)
\begin{equation}
  {{1}\over{2}}({\cal M}_{\rm s}^2-1)-\ln{\cal M}_{\rm s}=
{{\Delta\Phi}\over{c_{\rm s}^2}},
\end{equation}
where
\begin{equation}
  \Delta\Phi={{(\Omega_0H-2\Omega_{1{\rm s}}r\tan i)^2}\over{2(3\tan^2i-1)}}
\end{equation}
is the potential barrier to the outflow.  The vertical mass flux
density in the outflow is then
\begin{equation}
  \dot m_{\rm w}=\rho u_z={\cal M}_{\rm s}\rho_{\rm s} c_{\rm s}\cos i.
\end{equation}

\subsection{Non-dimensionalization}

We now rewrite the equations in a non-dimensional form suitable for
numerical analysis.  Given the local surface density $\Sigma$, the
angular velocity $\Omega_0$, and the constants appearing in the
constitutive relations, we identify
\begin{eqnarray}
  U_H&=&\Sigma^{(2+x)/(6+x-2y)}\Omega_0^{-(5-2y)/(6+x-2y)}
  \nonumber\\
  &&\times
  \left({{\mu m_{\rm H}}\over{k}}\right)^{-(4-y)/(6+x-2y)}
  \left({{16\sigma}\over{3C_\kappa}}\right)^{-1/(6+x-2y)}
\end{eqnarray}
as a natural unit for the semi-thickness of the disk.  Natural units
for other physical quantities follow according to
\begin{equation}
  U_\rho=\Sigma U_H^{-1},\qquad
  U_p=\Sigma\Omega_0^2U_H,
\end{equation}
\begin{equation}
  U_T=\Omega_0^2\left({{\mu m_{\rm H}}\over{k}}\right)U_H^2,\qquad
  U_F=\Omega_0U_HU_p,
\end{equation}
\begin{equation}
  U_B=(\mu_0U_p)^{1/2}.
\end{equation}
Note that the above expression for $U_H$ can be obtained from the
condition
\begin{equation}
  U_F={{16\sigma U_T^3}\over{3C_\kappa U_\rho^x U_T^y U_\rho}}
  {{U_T}\over{U_H}},
\end{equation}
which is a dimensional analysis of the definition (\ref{f}) of the
radiative flux.

There are two small dimensionless parameters in the problem,
\begin{equation}
  \epsilon={{U_H}\over{r}},\qquad
  \delta={{U_F}\over{\sigma U_T^4}}.
\end{equation}
Evidently $\epsilon$ is a characteristic measure of the angular
semi-thickness of the disk, while $\delta$ is an inverse measure of
the total optical thickness.

Non-dimensional variables are then introduced using the
transformations
\begin{equation}
  z=z_*\,U_H,\qquad
  H=H_*\,U_H,
\end{equation}
\begin{equation}
  \rho=\rho_*\,U_\rho,\qquad
  p=p_*\,U_p
\end{equation}
\begin{equation}
  T=T_*\,U_T,\qquad
  F=F_*\,U_F,
\end{equation}
\begin{equation}
  \nu=\nu_*\,\Omega_0U_H^2,\qquad
  \eta=\eta_*\,\Omega_0U_H^2,
\end{equation}
\begin{equation}
  u_r=u_{r*}\,\Omega_0U_H,\qquad
  \Omega_1=\Omega_{1*}\,\epsilon\Omega_0,
\end{equation}
\begin{equation}
  B_r=B_{r*}\,U_B,\qquad
  B_\phi=B_{\phi*}\,U_B,
\end{equation}
\begin{equation}
  B_z=B_{z*}\,U_B,\qquad
  {{\partial\psi_0}\over{\partial t}}=\dot\psi_*\,r\Omega_0U_HU_B.
\end{equation}
The transformed equations are
\begin{equation}
  -2\rho_*\Omega_{1*}=B_{z*}B_{r*}',
  \label{omega1}
\end{equation}
\begin{equation}
  {{1}\over{2}}\rho_*u_{r*}=B_{z*}B_{\phi*}'-
  {{3}\over{4}}\epsilon(1+2D_{\nu\Sigma})\rho_*\nu_*+
  {{3}\over{2}}\epsilon D_H(\rho_*\nu_*z_*)',
  \label{motion_phi*}
\end{equation}
\begin{equation}
  0=-\rho_*z_*-p_*'-B_{r*}B_{r*}'-B_{\phi*}B_{\phi*}',
\end{equation}
\begin{equation}
  \dot\psi_*+u_{r*}B_{z*}=
  \eta_*\left(\epsilon D_B B_{z*}-B_{r*}'\right),
  \label{induction_rz*}
\end{equation}
\begin{equation}
  0=-{{3}\over{2}}B_{r*}+B_{z*}\Omega_{1*}'+(\eta_*B_{\phi*}')',
  \label{o1p}
\end{equation}
\begin{equation}
  F_*'={{9}\over{4}}\rho_*\nu_*+
  \eta_*\left[(B_{r*}')^2+(B_{\phi*}')^2\right],
\end{equation}
\begin{equation}
  F_*=-{{T_*^{3-y}}\over{\rho_*^{1+x}}}T_*',
\end{equation}
\begin{equation}
  p_*=\rho_*T_*,
\end{equation}
where
\begin{equation}
  D_{\nu\Sigma}={{\partial\ln(\bar\nu\Sigma)}\over{\partial\ln r}},\qquad
  D_H={{\partial\ln H}\over{\partial\ln r}},\qquad
  D_B={{\partial\ln B_z}\over{\partial\ln r}}
\end{equation}
are three dimensionless parameters, and the prime denotes
differentiation with respect to $z_*$.

The dimensionless viscosity and magnetic diffusivity are given by
\begin{equation}
  \nu_*={\rm Pm}\,\eta_*=\alpha T_*.
\end{equation}

The mid-plane symmetry conditions are
\begin{equation}
  F_*(0)=B_{r*}(0)=B_{\phi*}(0)=0.
\end{equation}
The photospheric boundary conditions are
\begin{equation}
  \delta F_*(H_*)=[T_*(H_*)]^4
\end{equation}
and
\begin{equation}
  [\rho_*(H_*)]^{1+x}[T_*(H_*)]^{1+y}=
  {{1}\over{8}}\delta(1+x)[H_*-2\Omega_{1*}(H_*)\tan i],
\end{equation}
together with
\begin{equation}
  B_{r*}(H_*)=B_{r{\rm s}*},\qquad
  B_{\phi*}(H_*)=B_{\phi{\rm s}*}.\qquad
\end{equation}
Finally, the definition of $\Sigma$ implies the normalization condition
\begin{equation}
  2\int_0^{H_*}\rho_*\,dz_*=1.
\end{equation}
When the solution is found, the optical depth at the mid-plane can be
determined from
\begin{equation}
  \tau_{\rm c}={{2}\over{3}}+\int_0^H\kappa\rho\,dz
  ={{2}\over{3}}+{{16}\over{3\delta}}\int_0^{H_*}\rho_*^{1+x}T_*^y\,dz_*.
\end{equation}

The fact that the small parameter $\epsilon$ cannot be fully scaled
out of the equations indicates that a strictly self-consistent
thin-disk asymptotic solution cannot be obtained when all the physical
effects we have considered are included.  The $\epsilon$ terms in
equation (\ref{motion_phi*}) are expected to be important only in the
case of a weak magnetic field.  The $\epsilon$ term in equation
(\ref{induction_rz*}) may be important when the field is nearly
vertical.

\subsection{Stability considerations}

Under the fixed alpha hypothesis the magnitudes of the turbulent
viscosity and magnetic diffusivity are determined by the prescribed
value of $\alpha$.  A problem with this approach, as will be seen in
the next section, is that equilibrium solutions can then be obtained
that are magnetorotationally unstable even when the effect of the
turbulent diffusivity on stability is taken into account.  This
suggests that the model may be physically inconsistent for these
examples, because the `channel solution' (Hawley \& Balbus 1991) would
continue to grow exponentially.  Furthermore, there is a close
connection between the stability of the equilibria and properties such
as the shape of the field lines (Ogilvie 1998).  We will find
solutions that we believe to be unstable, in which the field lines
bend more than once, and which have other undesirable properties.

A resolution of these difficulties is suggested by the numerical
simulations of magnetorotational turbulence, which indicate that {\it
  the turbulence is much more vigorous in the presence of a mean
  poloidal magnetic field} (Hawley, Gammie, \& Balbus 1995), provided
that the field is not so strong as to suppress the instability.
Indeed, Stone et al. (1996) found it impossible to run a simulation in
a stratified disk with a net vertical field, while Miller \& Stone
(2000) found a dramatic difference between simulations with no net
field and those with a fairly weak uniform vertical field.

In an effort to understand and model this behavior, we propose the
following physical hypothesis for turbulent disks with a mean field:
{\it the value of $\alpha$ adjusts so that the equilibrium is
  marginally stable to the magnetorotational instability when the
  turbulent magnetic diffusivity is taken into account\/}.  We remark
that a similar situation occurs in stellar convection, where an
equivalent hypothesis can be used as a basis for the mixing-length
theory.  In a further example of this approach, Kippenhahn \& Thomas
(1978) modeled the outcome of a shear instability by imposing marginal
stability according to the Richardson criterion.

When the mean poloidal magnetic field is very weak, the instability
favors small length scales and only a small value of $\alpha$ is
required to suppress it.  For stronger fields, the required value of
$\alpha$ will be larger.  For even stronger fields with
$\beta\approx1$ on the mid-plane, the instability is suppressed or
nearly so even without any diffusivity, and $\alpha$ will be small or
zero.  This variation of $\alpha$ with the strength of the mean
poloidal field is qualitatively in agreement with that found in
numerical simulations (Hawley, Gammie, \& Balbus 1995; Brandenburg
1998), suggesting that our physical hypothesis is reasonable.  However, our
model will require this hypothesis to extend into a parameter range
that cannot be (or at least has not been) reproduced in numerical
simulations of stratified disks.

Although a magnetorotationally unstable disk may contain numerous
unstable modes, it has been argued by Ogilvie (1998) that, in ideal
MHD, the last mode to be stabilized is an axisymmetric mode with
vanishing radial wavenumber (i.e. $\partial/\partial r\sim r^{-1}$
rather than $\partial/\partial r\sim H^{-1}$), and is the first mode
of odd symmetry about the mid-plane.  Marginal stability of the
annulus is then imposed by locating the marginal stability condition
for this critical mode.  We conjecture that this remains true when
dissipation is included.

From a consideration of the linearized equations in the
limit of vanishing eigenfrequency, we find that such a mode should
involve only horizontal motions and satisfy the horizontal components
of the perturbed equation of motion,
\begin{equation}
  -2\rho\Omega_0\,\Delta u_\phi=
  {{B_z}\over{\mu_0}}{{\partial\,\Delta B_r}\over{\partial z}},
\end{equation}
\begin{equation}
  {{1}\over{2}}\rho\Omega_0\,\Delta u_r=
  {{B_z}\over{\mu_0}}{{\partial\,\Delta B_\phi}\over{\partial z}},
\end{equation}
and the perturbed induction equation,
\begin{equation}
  0=B_z{{\partial\,\Delta u_r}\over{\partial z}}+
  {{\partial}\over{\partial z}}
  \left(\eta{{\partial\,\Delta B_r}\over{\partial z}}\right),
\end{equation}
\begin{equation}
  0=-{{3}\over{2}}\Omega_0\,\Delta B_r+
  B_z{{\partial\,\Delta u_\phi}\over{\partial z}}+
  {{\partial}\over{\partial z}}
  \left(\eta{{\partial\,\Delta B_\phi}\over{\partial z}}\right),
\end{equation}
where $\Delta$ denotes a linearized Eulerian perturbation.  These
equations should be compared with the unperturbed equations
(\ref{motion_r})--(\ref{motion_phi}) and
(\ref{induction_rz})--(\ref{induction_phi}).  Note that this problem
is different from the marginal stability problem considered in
Paper~I, where the magnetic diffusivity was zero.

The critical mode is expected to be the first mode of odd symmetry
about the mid-plane (Ogilvie 1998).  This should satisfy the symmetry
conditions $\Delta u_r=\Delta u_\phi=0$ on the mid-plane, and the
photospheric boundary conditions $\Delta B_r=\Delta B_\phi=0$ (since
$B_{r{\rm s}}$ and $B_{\phi{\rm s}}$ are fixed in the local model).
For such a mode, the linearized equations may be combined into the
second-order ODE
\begin{equation}
  {{\partial}\over{\partial z}}
  \left[\left({{B_z^2}\over{\mu_0\rho}}+
  {{\mu_0\rho\eta^2\Omega_0^2}\over{B_z^2}}\right)
  {{\partial\,\Delta B_r}\over{\partial z}}\right]+
  3\Omega_0^2\,\Delta B_r=0,
\end{equation}
or, in dimensionless form,
\begin{equation}
  \left[\left({{B_{z*}^2}\over{\rho_*}}+
  {{\rho_*\eta_*^2}\over{B_{z*}^2}}\right)\Delta B_{r*}'\right]'+
  3\,\Delta B_{r*}=0,
\end{equation}
with boundary conditions
\begin{equation}
  \Delta B_{r*}'(0)=\Delta B_{r*}(H_*)=0.
\end{equation}

We are now in a position to determine the value of $\alpha$ as
follows.  For any given value of $\alpha$ and suitable values of the
other parameters, we might expect to find an equilibrium solution.
However, the equilibrium will not in general possess a marginal mode
satisfying the correct boundary conditions.  By integrating the
equations for a marginal mode simultaneously with those defining the
equilibrium, we aim to tune $\alpha$ until a marginal mode is
obtained.  If this mode is the first mode of odd symmetry, we then
believe we have determined the value of $\alpha$ corresponding to a
marginally stable equilibrium.

A crude estimate of this may be obtained by approximating the second
derivative with respect to $z_*$ as $-H_*^{-2}$.  This leads to the
estimate
\begin{equation}
  \eta_*^2\approx{{B_{z*}^2}\over{\rho_*}}
  \left(3H_*^2-{{B_{z*}^2}\over{\rho_*}}\right),
  \label{crude}
\end{equation}
which has the properties described above.

It might be argued that our marginal stability condition is too strong
because the presence of one or two unstable modes of relatively long
radial wavelength might not contribute significantly to turbulent
transport.  However, if the critical mode is not stabilized, it will
lead to the growth of the fundamental `channel solution', which is
very efficient in transporting angular momentum and probably
hydrodynamically unstable in three dimensions, leading to enhanced
turbulence.  Admittedly, the description of the disk under these
circumstances is a matter of some uncertainty.

\section{Numerical investigation}

\subsection{Numerical method}

The dimensionless ODEs are integrated from the photosphere $z_*=H_*$
to the mid-plane $z_*=0$.  The dependent variables are $p_*$, $T_*$,
$F_*$, $\Omega_{1*}$, $B_{r*}$, and $B_{\phi*}$.  The values of $H_*$,
$F_*(H_*)$, $\Omega_{1*}(H_*)$, and $\dot\psi_*$ are guessed and then
adjusted by Newton-Raphson iteration to match the symmetry conditions
on the mid-plane.  In the marginal stability model, the equations for
the marginal mode are integrated simultaneously.  An arbitrary
normalization $\Delta B_{r*}'(H_*)=1$ of the linear problem is
applied, and the value of $\alpha$ is also guessed and adjusted by
Newton-Raphson iteration to match the symmetry condition for the mode
on the mid-plane.

To obtain the derivatives $B_{\phi*}'$ and $\Omega_{1*}'$ some algebra
is required.  Eliminating $u_{r*}$, $p_*'$, and $B_{r*}'$ we obtain
\begin{eqnarray}
  &&B_{\phi*}'=
  {{\rho_*}\over{2{\rm Pm}\,B_{z*}^2
  (2B_{z*}-3\epsilon D_H\alpha B_{\phi*}z_*)}}\nonumber\\
  &&\times\Bigg\{-2{\rm Pm}\,B_{z*}\dot\psi_*+4\alpha p_*\Omega_{1*}
  \nonumber\\
  &&\qquad
  +\epsilon\left[2D_B+3{\rm Pm}\,(1+2D_{\nu\Sigma}-2D_H)\right]
  \alpha B_{z*}^2T_*
  \nonumber\\
  &&\qquad+6{\rm Pm}\,\epsilon D_H\alpha B_{z*}
  \left(B_{z*}z_*^2-
  2B_{r*}\Omega_{1*}z_*\right)\Bigg\}.
\end{eqnarray}
By multiplying this expression by $\eta_*$ and differentiating, we
find $\Omega_{1*}'$ from equation (\ref{o1p}).

\subsection{Unmagnetized solution}

In the absence of a mean magnetic field, a solution is obtained by
omitting the induction equation and setting $\bB={\bf0}$ elsewhere.
This also implies $\Omega_1=0$.  The form of the solution for $p_*$,
$T_*$, and $F_*$ depends only on the dimensionless parameter $\delta$
(for a given opacity law), although the variables also have power-law
scalings with $\alpha$.  If $u_{r*}$ is required, there are further
dependences on $D_{\nu\Sigma}$ and $D_H$, and the value of $u_{r*}$ is
proportional to $\epsilon$ as a consequence of the scalings we have
adopted for the general problem.

We focus on the case of Thomson scattering opacity ($x=y=0$,
$C_\kappa\approx0.33\,{\rm cm}^2\,{\rm g}^{-1}$).  In a steady disk,
far from the inner edge, we may then set $D_{\nu\Sigma}=0$ and
$D_H=21/20$ (Shakura \& Sunyaev 1973).  We also take $\alpha=0.1$.
For the purposes of illustration, we consider a location at 1000
Schwarzschild radii from a black hole of mass $10M_\odot$, i.e.
$r=2.95\times10^9\,{\rm cm}$.  For a surface density
$\Sigma=10^4\,{\rm g}\,{\rm cm}^{-2}$, and assuming $\mu=0.6$, we find
illustrative values $U_H=6.34\times10^7\,{\rm cm}$,
$U_\rho=1.58\times10^{-4}\,{\rm g}\,{\rm cm}^{-3}$,
$U_p=3.27\times10^{10}\,{\rm dyn}$, $U_T=1.50\times10^6\,{\rm K}$,
$U_F=4.70\times10^{17}\,{\rm erg}\,{\rm cm}^{-2}\,{\rm s}^{-1}$, and
$U_B=6.41\times10^5\,{\rm G}$.  Then $\epsilon=2.15\times10^{-2}$ and
$\delta=1.62\times10^{-3}$.

The numerically determined unmagnetized solution is shown in Fig.~1.
It has a dimensionless photospheric height of $H_*=1.70$ and an
optical depth at the mid-plane of $\tau_{\rm
  c}={\textstyle{{2}\over{3}}}+{\textstyle{{8}\over{3}}}\delta^{-1}$.
Thus our illustrative values correspond to $H/r=0.0364$ and $\tau_{\rm
  c}=1650$, representing a geometrically thin and optically thick
disk.  The illustrative accretion rate is $\dot
M=3.55\times10^{18}\,{\rm g}\,{\rm s}^{-1}$, which is less than the
Eddington accretion rate $1.69\times10^{19}\,{\rm g}\,{\rm s}^{-1}$
for an accretion efficiency of $0.1$.

The profile of radial velocity is of particular interest, since this
will affect the advection of the magnetic field.  The radial velocity
is {\it positive\/} on the mid-plane and becomes negative at larger
$z$, the density-weighted average being of course negative.  This
result has been found previously by several authors (e.g.\ Urpin 1984;
Kley \& Lin 1992, where further analysis of this phenomenon may be
found).  In the present model it results from the vertical profile of
the viscosity, specifically from the $\partial\ln H/\partial\ln r$
term in equation (\ref{ur}).

\subsection{Magnetized solutions under the fixed alpha hypothesis}

We first adopt the fixed alpha hypothesis and search for magnetized
solutions, taking parameter values $\alpha=0.1$, $B_{\phi{\rm s}*}=0$,
${\rm Pm}=1$, $D_B=0$ in addition to $x=y=0$, $D_{\nu\Sigma}=0$,
$D_H=21/20$, and the illustrative values for $\epsilon$ and $\delta$.
There are two limits in which physically acceptable solutions are
easily obtained over a wide range of inclination angles: very weak
fields, and strong fields.  We illustrate two such cases, with
$i=45\degr$, in Figs~2 and~3.

In the very weakly magnetized solutions, the magnetic field behaves
kinematically and the equilibrium structure of the unmagnetized
solution is not significantly distorted (e.g.\ the radial velocity
profile is almost indistinguishable from that in the unmagnetized
solution).  The poloidal field lines bend smoothly according to a
balance between advection and diffusion.  Isorotation is not enforced
(there is only a minuscule deviation from Keplerian rotation) and the
field is strongly wound up ($|B_\phi|>|B_z|$) in the example shown,
with $i=45\degr$.  The poloidal field lines continue to bend as the
photosphere is approached.  This is permitted because $\beta$ is very
large throughout.  Our atmospheric model, which assumed $\beta<1$, is
not relevant here and certainly the outflow solution should be
disregarded.  It is not clear how to treat the atmosphere in a case
such a this, because it is not understood whether the radial flow and
magnetic diffusivity continue above the photosphere.

By setting $B_{\phi{\rm s}}=0$, we have ensured that the radial mass
flux is caused by viscosity alone (cf. eq.~[\ref{am}]).  As expected,
the radial velocity causes the flux to migrate inwards for almost
vertical fields, but outwards for larger inclination angles.  With
${\rm Pm}=1$ and $D_B=0$, one finds the intermediate case
$\dot\psi_*=0$ at $i=0.485=27.8\degr$.  This is much larger than
$H/r=0.0364$, indicating that our concerns expressed in Section~3 were
well founded.  Alternatively, one can prevent the flux from being
expelled (i.e.\ achieve $\dot\psi_*=0$) when $i=45\degr$ by increasing
${\rm Pm}$ from $1$ to $1.90$. Increasing $D_B$ does not change
$\dot\psi_*$ significantly.

For strong fields, we recover solutions very similar to those we
obtained in Paper~I.  In the example shown in Fig.~3, the disk is
significantly compressed by the Lorentz force (compare the plot of
$p_*$ with the unmagnetized case), and is hotter (cf.\ $T_*$) and more
luminous (cf.\ $F_*$) than its unmagnetized counterpart.  Most of the
bending of the poloidal field lines occurs near the mid-plane where
$\beta$ is close to unity.  The field enforces isorotation, resulting
in sub-Keplerian rotation, and very little toroidal field is produced.

A feature of the solution is that the radial velocity, although
everywhere subsonic, is rather large and non-uniform in direction.
The reason for the profile of $u_{r*}$ is that the flux is being
expelled rapidly ($\dot\psi_*=-0.551$) by turbulent diffusion.  In the
upper layers, where $\beta$ is small, the fluid must follow the field
and therefore flows rapidly outwards.  To achieve the net accretion
rate imposed by the vertically integrated viscous stress, a strong
inflow is then required close to the mid-plane.  Such a profile might
even be considered advantageous for jet launching, since the outflow
itself must involve a positive radial velocity.

With $B_{z*}=2$, as in this example, and assuming ${\rm Pm}=1$, zero
flux migration is found at $i=0.0346=1.98\degr$.  This is now close to
$H/r=0.0364$ as expected from simple arguments, probably because the
field-line bending occurs close to the mid-plane.

On the other hand, these considerations may be inconsistent because
the example in Fig.~3 is magnetorotationally stable and therefore
unlikely to be turbulent.  Of course, if there is no turbulent
magnetic diffusivity, the origin of the viscosity should also be
questioned.

For a range of intermediate field strengths, the situation is more
complicated.  Solutions are not found in which the field lines bend in
the simple way seen in Figs~2 and~3.  Instead, branches of solutions
appear in which the field lines bend several times as they pass
through the disk.  An example is shown in Fig.~4.

According to Ogilvie (1998, Theorem~1) such a solution would be
magnetorotationally unstable in ideal MHD.  The multiple bending is
closely related to the appearance of the `channel solution' which is
the first stage of the magnetorotational instability for a vertical
field.  We conjecture that such multiple-bending solutions are also
unstable when the turbulent magnetic diffusivity is taken into account
(in both the equilibrium and perturbation equations), although this
has not been proven.  If correct, this would imply that solutions of
this type are physically inconsistent and ought to be disregarded,
which in turn would suggest that no steady solution is possible for
these intermediate field strengths.

\textit{However, our alternative proposal, the marginal stability
hypothesis, allows for a possible solution of this difficulty.}

\subsection{Magnetized solutions under the marginal stability hypothesis}

Under the marginal stability hypothesis, the strength of the
turbulence (quantified through the parameter $\alpha$) is just
sufficient to bring the equilibrium to marginal magnetorotational
stability.  This enables us to find single-bending solutions in a
continuous range of field strengths from very small values up to the
strength at which the instability is suppressed even without a
turbulent diffusivity.  We adopt the same parameters as in the
previous section, except that $\alpha$ is now to be determined as part
of the solution.  Fig.~5 shows how the required value of $\alpha$
varies with $B_{z*}$ for solutions with vertical fields ($i=0$).

This has the general form expected from the crude estimate, equation
(\ref{crude}).  However, it is somewhat worrying that values of
$\alpha$ in excess of unity may be required.  It is often argued that
$\alpha$ should not exceed unity because the turbulence would then
have to be supersonic, or to have a correlation length greater than
the disk thickness (e.g.\ Pringle 1981).  It is not clear whether
these constraints necessarily apply to magnetorotational turbulence,
which is dominated by anisotropic Maxwell stresses whose correlation
length in the azimuthal direction could exceed $H$ (e.g.\ Armitage
1998), but which could be limited instead by magnetic buoyancy.
Moreover, the effective transport coefficients may themselves be
anisotropic in the presence of a strong mean field (e.g.\ Matthaeus
et~al.\ 1998) and the magnetic Prandtl number may also differ from
unity.  Indeed, it is not certain whether the effects of the
turbulence can be described adequately in terms of an effective
diffusivity; it may be that coherent magnetic structures are formed.
Nevertheless, the question remains as to whether the turbulence can
truly reach a level sufficient to achieve marginal stability in the
presence of a significant mean field.  The numerical simulations by
Hawley et~al.\ (1995) are in qualitative agreement with Fig.~5 but
suggest that, when the field strength is just below the stability
boundary for a given computational domain, the channel solution may
continue to grow without degenerating into turbulence.  However,
simulations of stratified models with fairly strong mean vertical
fields do not appear to have been successful.

An example solution calculated under the marginal stability hypothesis
is shown in Fig.~6.  This has $\alpha=3.81$ and should be contrasted
with the solution in Fig.~4 which has the same vertical field strength
but a fixed $\alpha=0.1$.  The marginally stable solution has field
lines with a single bend which become straight not far below the
photosphere.  The pressure and temperature decline smoothly and
monotonically with increasing height.  The field is not significantly
wound up ($|B_\phi|<|B_z|$).

\subsection{Mass loss rates for jet-launching disks}

The principal aim of this paper was to determine how the mass loss
rate in the outflow, $\dot m_{\rm w}$, varies with the strength and
inclination of the magnetic field.  In Fig.~7 we show the result of
this calculation under the fixed alpha hypothesis with $\alpha=0.1$
and other parameters as given in Section~5.3.  Note that the
reciprocal of $\dot m_{\rm w}/(\Sigma\Omega)$ is approximately the
number of orbits in which the disk would be evaporated if not
replenished by the accretion flow.  In Fig.~8 we plot the quantity
$4\pi r^2\dot m_{\rm w}/\dot M$ as a function of $B_{z*}$ for
solutions with $i=45\degr$.  This alternative dimensionless measure of
the outflow rate is the local mass loss rate per unit logarithmic
interval in radius, divided by the accretion rate.

The outcome is as we found in Paper~I.  The solutions shown are
magnetorotationally stable.  Close to the edge of the solution
manifold, the outflow is very vigorous, but if the field strength is
increased by only a factor of two above the stability boundary, the
outflow is suppressed by twenty orders of magnitude or so.  As
explained in Paper~I, this happens because the disk becomes
significantly sub-Keplerian (cf.\ Fig.~3) and the outflow experiences
a large potential barrier (whose strength is roughly proportional to
$B_z^4$).  This means that, in the absence of additional heating,
external irradiation, or other driving mechanisms, outflows are
suppressed from strongly magnetized disks.  For a fixed field
strength, the outflow is maximized at an intermediate inclination
angle of 40--$50\degr$.

The corresponding results for the marginal stability hypothesis are
shown in Figs~9 and~10.  Note that this is a complementary region of
parameter space corresponding to turbulent disks.  The behavior of the
mass loss rate is now completely different and perhaps more intuitive:
it increases monotonically with increasing $B_{z*}$ and with
increasing~$i$.  However, the solution branch terminates before
excessive mass loss rates are achieved.  This suggests that a large
potential barrier is not incurred for turbulent disks and that they
are more promising for jet launching.  The potential barrier is smaller
because the magnetic field is weaker, resulting in a smaller Lorentz
force and a smaller deviation from Keplerian rotation.

These two sets of results were obtained under different physical
assumptions.  Nevertheless, they both suggest that efficient
jet-launching solutions are found in a limited range of field
strengths, and in a limited range of inclination angles in excess of
$30\degr$.  In both cases there are difficulties in interpreting the
solutions.  The more strongly magnetized solutions obtained under the
fixed alpha hypothesis are magnetorotationally stable, and the origin
of the dissipation in the disk remains unclear.  The more weakly
magnetized solutions obtained under the marginal stability hypothesis
typically require values of $\alpha$ in excess of unity to bring the
equilibrium to marginal magnetorotational stability.

\section{Discussion}

We have developed a model of the local vertical structure of
magnetized accretion disks that launch magnetocentrifugal outflows.
Given certain assumptions concerning the dissipative processes
(turbulent viscosity and magnetic diffusivity) in the disk, we have
shown that it is possible to compute the mass loss rate in the outflow
as a function of the surface density and the strength and inclination
of the poloidal magnetic field.  The net accretion rates of mass and
magnetic flux are also determined.  This information is precisely
complementary to that obtained from numerical simulations of the
subsequent acceleration and collimation of a `cold' outflow (e.g.\ 
Krasnopolsky et~al.\ 1999).

The following result appears to be quite robust.  For disks in which
the mean poloidal magnetic field is sufficiently strong to stabilize
the equilibrium against the magnetorotational instability, we find
that the mass loss rate decreases extremely rapidly with increasing
field strength, and is maximized at an inclination angle of
40--$50\degr$.  For turbulent disks with weaker mean fields, we find
that the mass loss rate increases monotonically with increasing
strength and inclination of the field, but the solution branch
terminates before excessive mass loss rates are achieved.  This
suggests that turbulent disks with moderate mean fields are more
promising for jet launching, but there may be situations in which a
steady solution is impossible.

For each solution we have determined the net rate of flux migration,
which depends on a competition between inward dragging by the
accretion flow, and outward transport due to turbulent diffusion.  The
results depend on the effective magnetic Prandtl number of the
turbulence, which has never been measured.  However, we find that
inward migration is more likely to occur in the case of a weak mean
field, which bends at greater heights in the disk.  In this case,
inward dragging may be many times more effective than estimated in
previous studies (Lubow et~al.\ 1994a), with the result that a net
inward migration of flux may occur even when the inclination angle is
sufficient for jet launching, provided that the disk is not very thin.
This issue requires further analysis and a better understanding of the
kinematic behavior of the magnetic field.  For stronger fields, we
find fairly good agreement with previous studies, and the
jet-launching configurations are much more difficult to maintain
against dissipation.  We speculate that this may lead naturally to a
situation in which the flux is regulated to a value suitable for jet
launching.  However, we have not modeled the possible effect of an
internal dynamo in the disk.  In addition, it is quite probable that
instabilities will arise when the global coupling between the outflow
and the flux evolution is considered (Lubow, Papaloizou, \& Pringle
1994b; Cao \& Spruit, in preparation).

Several other authors have considered the vertical structure of
magnetized disks and the problem of the disk-jet connection.  Our
formulation is distinctive in that we have shown how to calculate the
mass loss rate and the accretion rates of mass and magnetic flux at
any radius from a knowledge of the local surface density and the
strength and inclination of the magnetic field.  When coupled with a
numerical simulation of the outflow beyond the sonic point, this would
provide a closed evolutionary scheme for a time-dependent magnetized
accretion disk with an outflow.  It is crucial that we allowed the
external magnetic torque on the disk to be determined by the exterior
outflow solution rather than the interior solution of the vertical
disk structure.

There are further differences in the detail of our approach.  Wardle
\& K\"onigl (1993) discussed many of the issues that we have
addressed, but we differ from them in considering an optically thick
disk with turbulent transport and energy dissipation, as opposed to an
isothermal, inviscid disk with ambipolar diffusion.  We are also able
to calculate the net rate of flux accretion rather than specifying it
as a parameter.  Similar comparisons may be made with the analysis of
Li (1995).  The model of Campbell (1999) shares some features with the
present paper, but his outflow solution treats the magnetic field
lines as parabolae throughout the disc and the transonic region, while
our solutions (e.g.\ Fig.~6) indicate that the field lines become
straight in the atmosphere, provided that the plasma beta is less than
unity.  If $\beta>1$ in the atmosphere, the magnetocentrifugal
mechanism does not work.

Interesting comparisons may be made with the steady, self-similar
model of Casse \& Ferreira (2000).  In common with those authors, we
find that values of $\alpha$ exceeding unity, and/or anisotropic
effective turbulent diffusion coefficients, may be helpful or
necessary in obtaining plausible solutions for efficient jet-launching
disks.

Admittedly, our model has some limitations.  Our simple modeling of
the effects of the turbulence is consistent with the very limited
information available from numerical simulations, although we have not
included a dynamo $\alpha$-effect.  We have assumed that the
atmosphere is magnetically dominated so that the poloidal field lines
act as rigid channels for the outflow.  This constraint could be
relaxed in future work at the expense of introducing considerable
complications in solving for the atmospheric flow, and in matching to
the exterior solution.

Future work should aim at obtaining a better understanding of the
interaction of the mean magnetic field with the turbulence.  Numerical
simulations might be used to measure the effective magnetic
diffusivity tensor and how it depends on the strength of the field.  A
crucial issue is whether, when the mean field is a significant
fraction of the value required for magnetorotational stability, the
`channel solution' of the instability persists or degenerates into
turbulence.

Ideally the whole problem that we have defined, including the correct
boundary conditions, would be solved using a numerical simulation.  We
have included some non-local effects (Section~4.4) but most of the
relevant terms are included in the shearing-box local accretion disk
model (e.g.\ Stone et~al.\ 1996).  It may be difficult, however, to
resolve the disk adequately from the mid-plane to the sonic point,
although the techniques used by Miller \& Stone (2000) should be
helpful in achieving this.

\acknowledgments

GIO acknowledges support from the STScI visitor program and from Clare
College, Cambridge.  ML acknowledges support from NASA Grants
NAG5-6857 and GO-07378.  We thank Jim Pringle and Henk Spruit for
helpful discussions.

\begin{figure}
  \centerline{\epsfbox{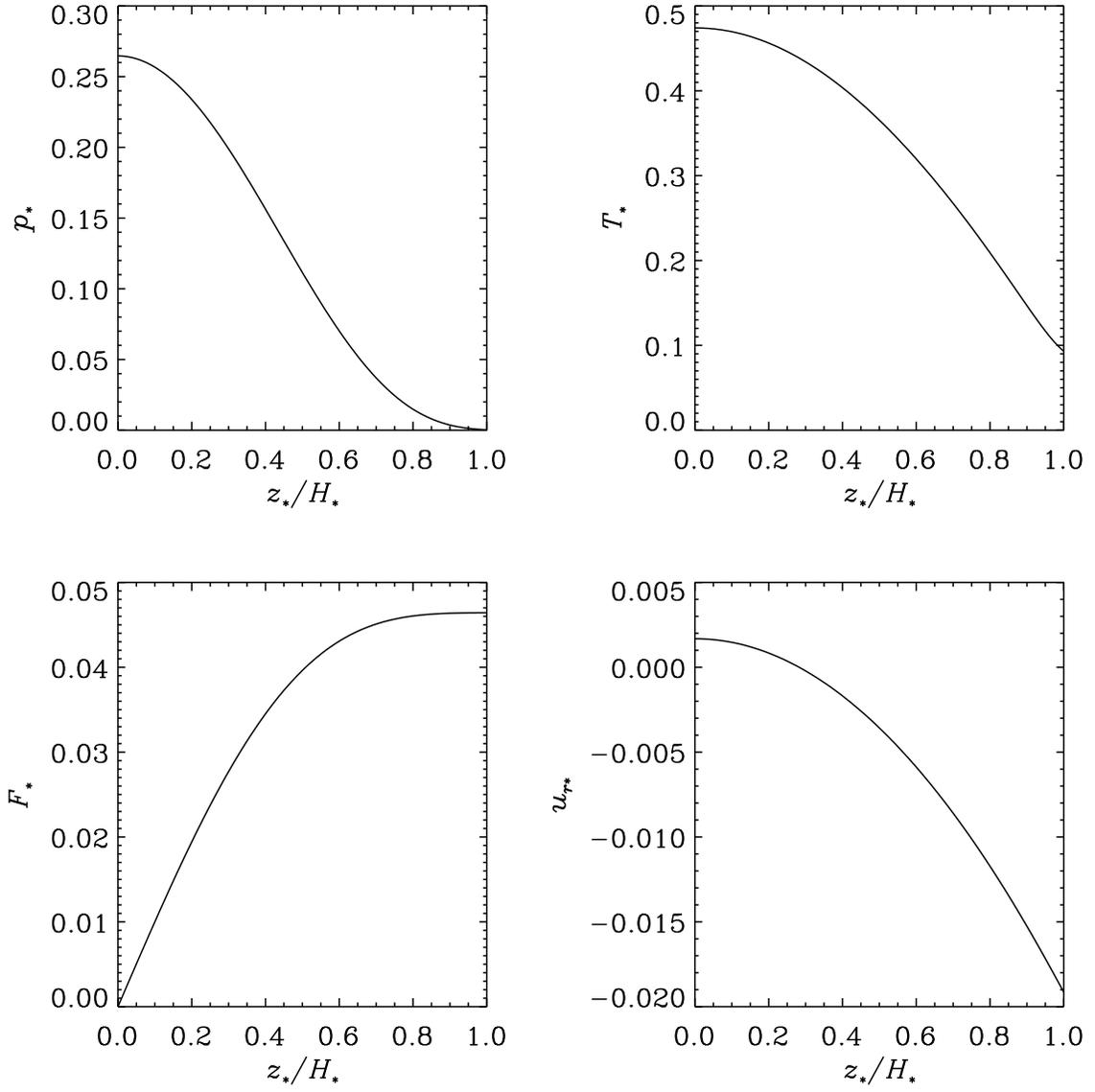}}
  \caption{Profiles of pressure, temperature, radiative energy flux, and
    radial velocity for an unmagnetized model with Thomson opacity.
    Note that in this case the radial velocity is directed outwards on
    the mid-plane and becomes inward at greater height.}
\end{figure}

\begin{figure}
  \centerline{\epsfbox{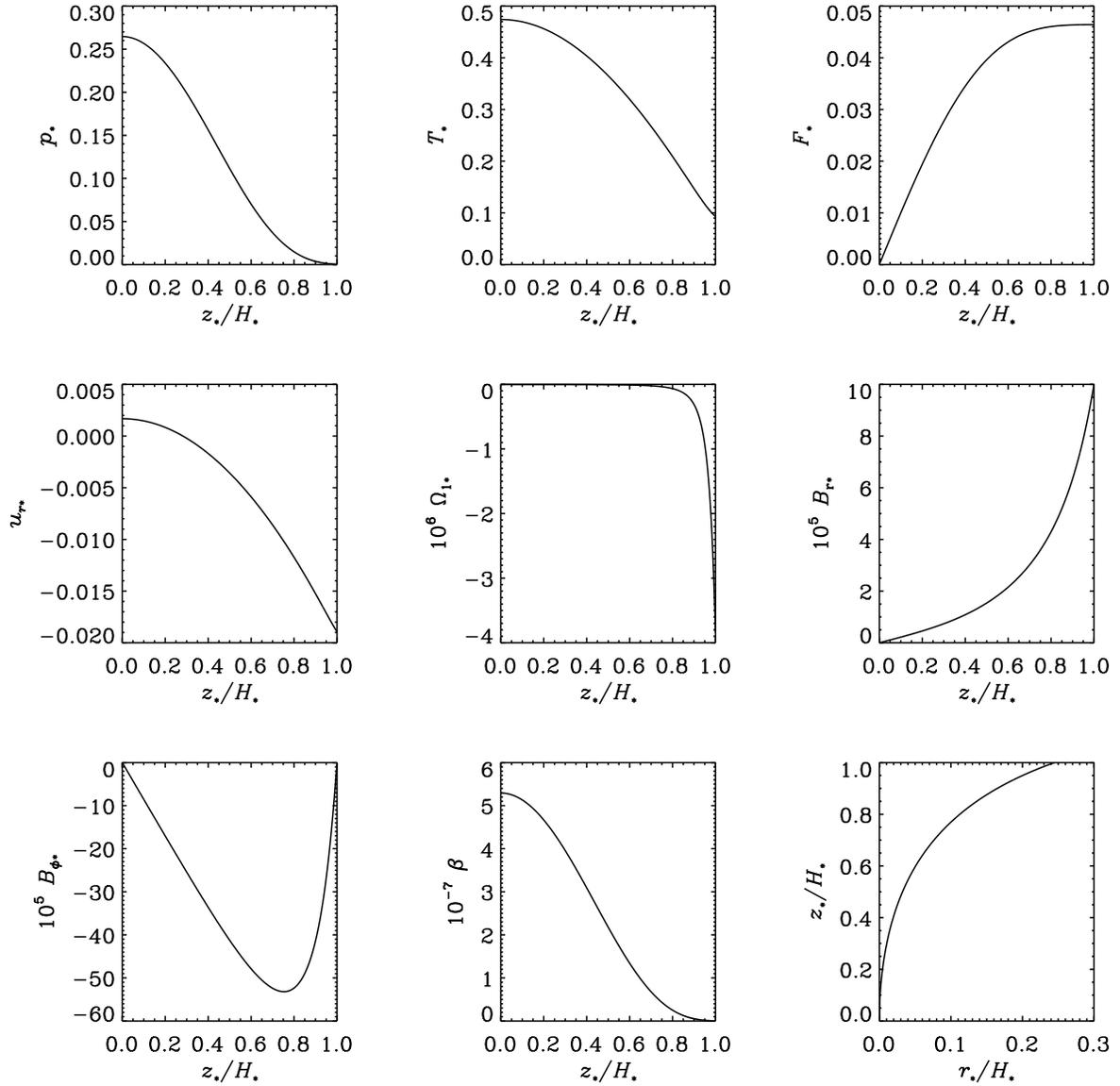}}
  \caption{Profiles of pressure, temperature, radiative energy flux,
    radial velocity, deviation from Keplerian angular velocity, radial
    magnetic field, toroidal magnetic field, and poloidal plasma beta,
    and shape of the poloidal field lines, for a very weakly
    magnetized model with $B_{z*}=10^{-4}$ (i.e. $B_z=64.1\,{\rm G}$
    for our illustrative parameters) and $i=45\degr$.}
\end{figure}

\begin{figure}
  \centerline{\epsfbox{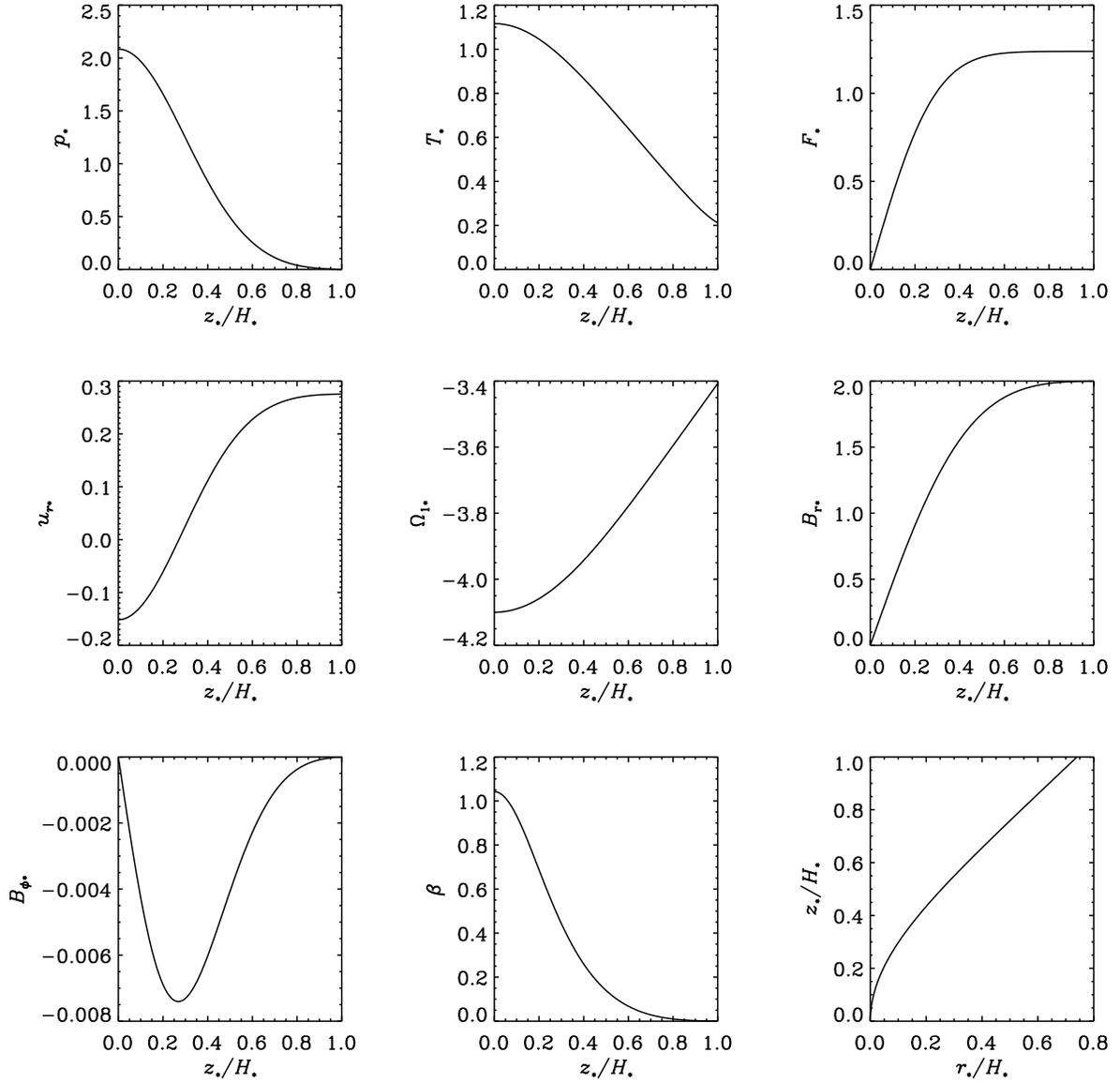}}
  \caption{A strongly magnetized model with $B_{z*}=2$ (i.e.
    $B_z=1.28\times10^6\,{\rm G}$ for our illustrative parameters) and
    $i=45\degr$.}
\end{figure}

\begin{figure}
  \centerline{\epsfbox{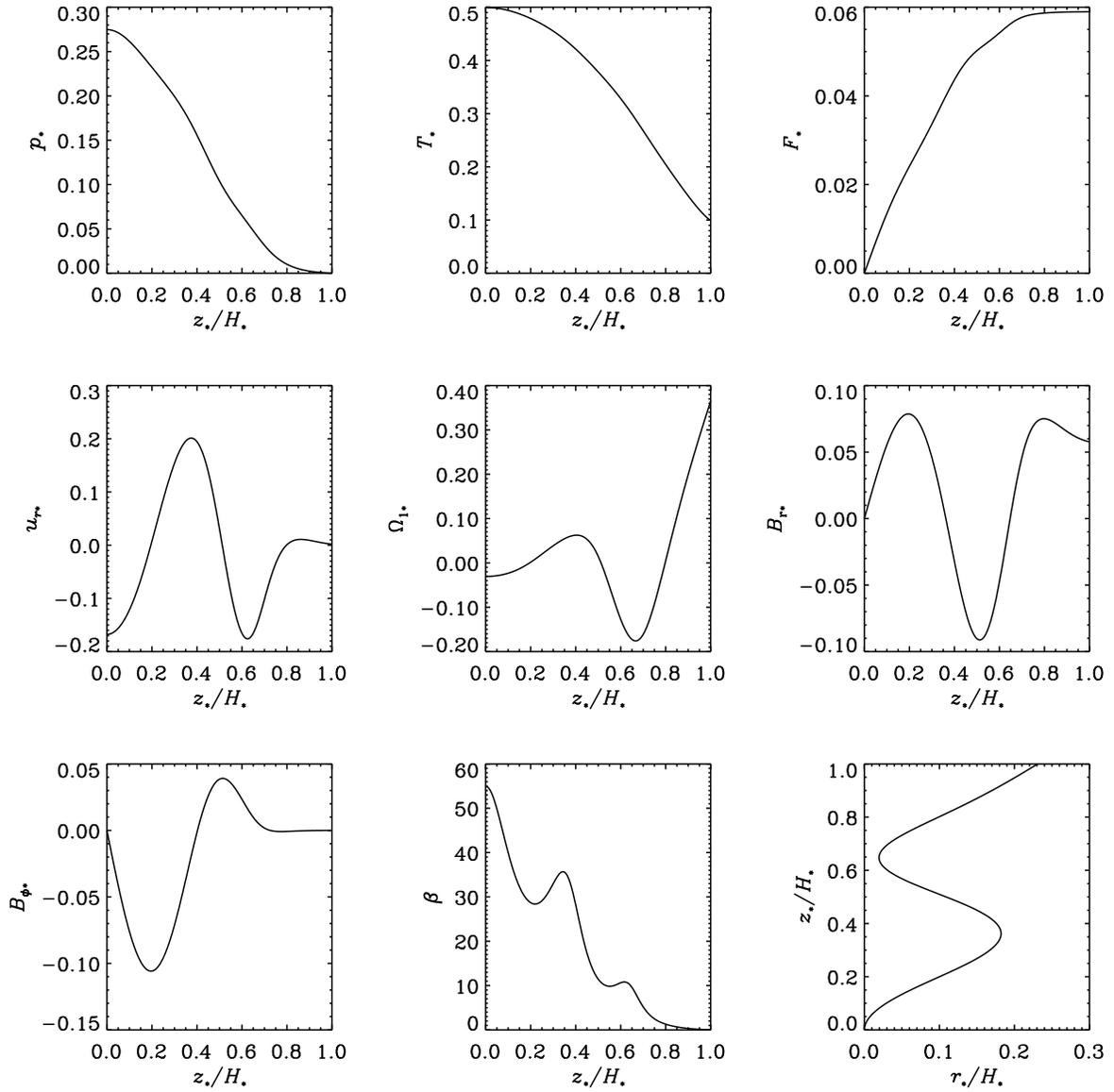}}
  \caption{A peculiar model with $B_{z*}=0.1$ (i.e.
    $B_z=6.41\times10^4\,{\rm G}$ for our illustrative parameters) and
    $i=30\degr$.  Note the multiple bending of the poloidal field
    lines, which probably indicates a magnetorotationally unstable
    configuration, and other undesirable characteristics.}
\end{figure}

\begin{figure}
  \centerline{\epsfysize=8cm\epsfbox{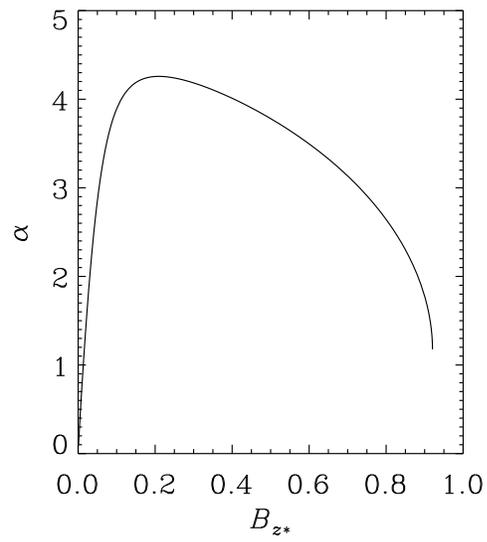}}
  \caption{Variation of the viscosity parameter $\alpha$ with the mean
    vertical field strength $B_{z*}$ under the marginal stability
    hypothesis, for solutions with vertical fields ($i=0$).}
\end{figure}

\begin{figure}
  \centerline{\epsfbox{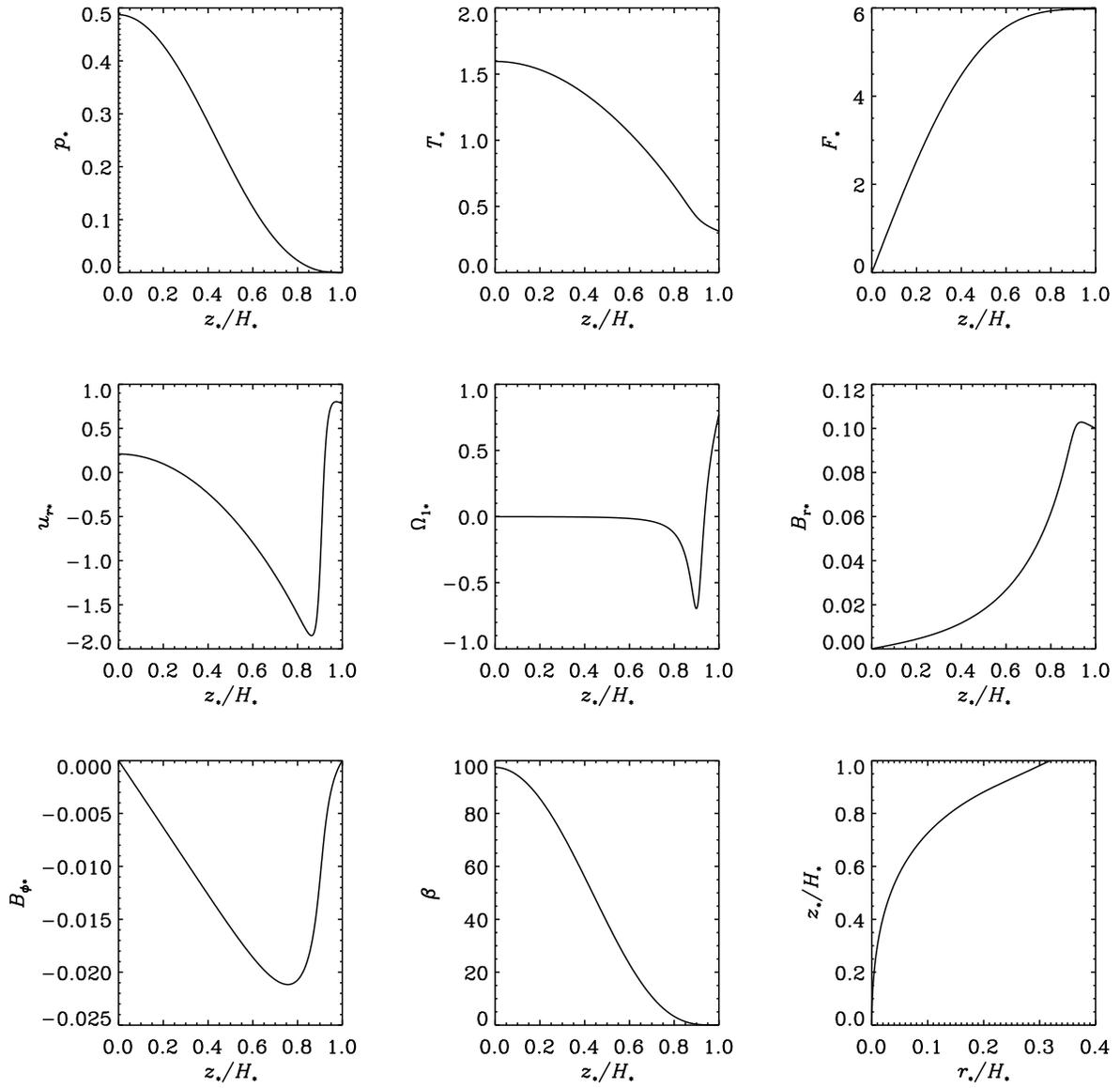}}
  \caption{A marginally stable model with $B_{z*}=0.01$ and $i=45\degr$.}
\end{figure}

\begin{figure}
  \centerline{\epsfysize=8cm\epsfbox{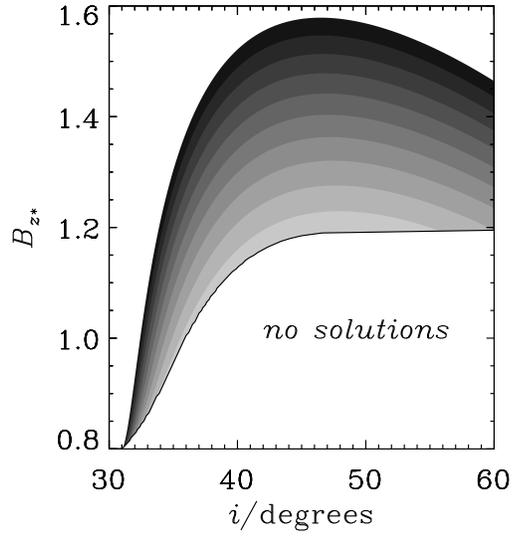}}
  \caption{Mass loss rate as a function of the strength and inclination of
    the magnetic field, under the fixed alpha hypothesis.  Contours of
    $\log_{10}(\dot m_{\rm w}/\Sigma\Omega)$ are plotted, from $-14$
    (darkest) to $-4$ (lightest) with unit spacing.  The solutions
    continue to larger values of $B_{z*}$, but do not extend below the
    solid line.}
\end{figure}

\begin{figure}
  \centerline{\epsfysize=8cm\epsfbox{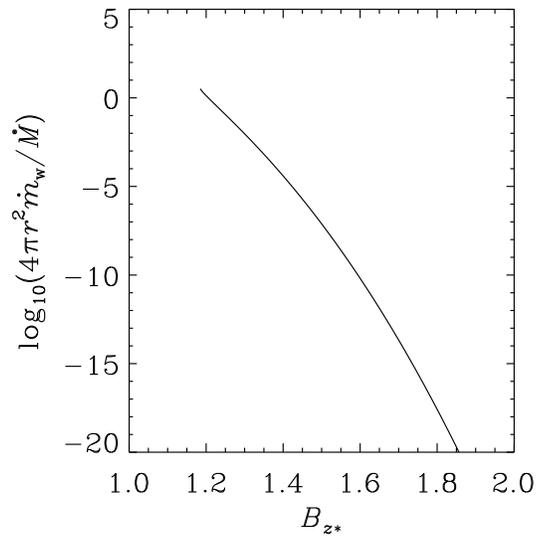}}
  \caption{Ratio of the mass loss rate to the accretion rate as a function
    of the strength of the magnetic field for equilibria with
    $i=45\degr$, under the fixed alpha hypothesis.  The same solutions
    appear in Fig.~7.}
\end{figure}

\begin{figure}
  \centerline{\epsfysize=8cm\epsfbox{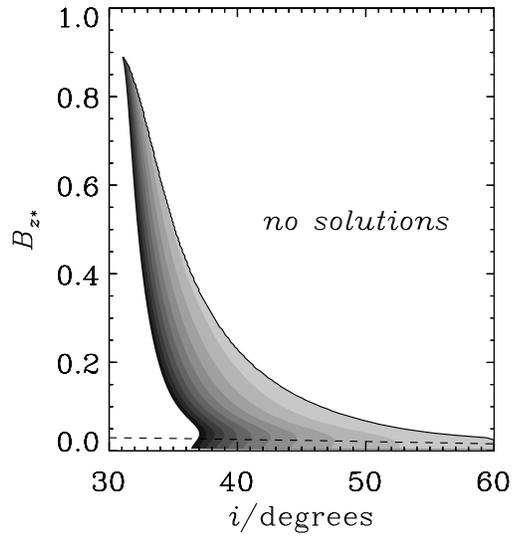}}
  \caption{Mass loss rate under the marginal stability hypothesis.
    Contours of $\log_{10}(\dot m_{\rm w}/\Sigma\Omega)$ are plotted,
    from $-14$ (darkest) to $-4$ (lightest).  The solutions continue
    to smaller values of $i$, but do not extend to the right of the
    solid line.  Below the dashed line, the atmosphere is not
    magnetically dominated ($\beta>1$ at the photosphere) and the
    outflow solution should not be trusted.}
\end{figure}

\begin{figure}
  \centerline{\epsfysize=8cm\epsfbox{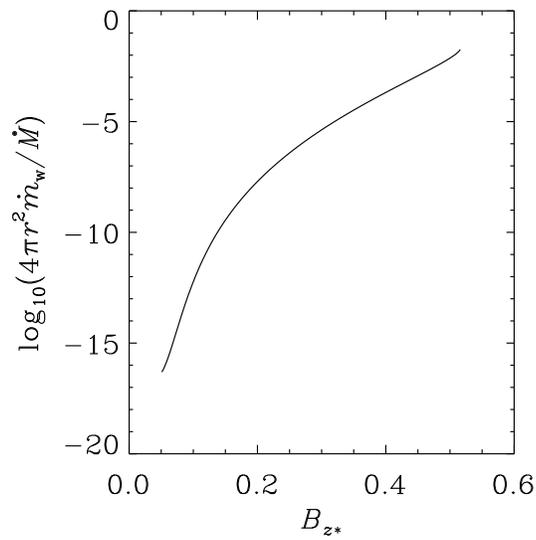}}
  \caption{Ratio of the mass loss rate to the accretion rate as a function
    of the strength of the magnetic field for equilibria with
    $i=35\degr$, under the marginal stability hypothesis.  The same
    solutions appear in Fig.~9.}
\end{figure}

\end{document}